\documentclass[showpacs,showkeys]{revtex4}
\usepackage{graphicx}

\begin{document}

\title{Nuclear effects in neutrino induced reactions}

\author{M. J. Vicente Vacas, L. Alvarez-Ruso, L. S. Geng}

\affiliation{Departamento de F\'{\i}sica Te\'orica and IFIC, 
Universidad de Valencia-CSIC,\\
46100 Burjassot (Valencia), Spain}

\author{J. Nieves, M. Valverde}

\affiliation{Departmento de F\'{\i}sica At\'omica Molecular y Nuclear, Universidad
de Granada,\\
18071 Granada, Spain}

\author{S. Hirenzaki}

\affiliation{Department of Physics, Nara Women's University, Nara, 630-8506, Japan}

\begin{abstract}
We discuss the relevance of nuclear medium effects in the analysis of  some low and
medium energy neutrino reactions of current interest.  In particular, we study the
Quasi-Elastic (QE) process, where RPA correlations and Final State Interactions (FSI)
are shown to play a crucial role. We have also investigated the $\nu$ induced
coherent pion production. We find a strong reduction of the cross section due
to the distortion of the pion wave function and the modification of the production
mechanisms in the nucleus. The sensitivity of the results to the axial $N\Delta$ 
coupling $C_5^A(0)$ has been also investigated. 
\end{abstract}

\maketitle

\section{Introduction}	

The analysis of the new generation of neutrino oscillation experiments requires a good knowledge of
the neutrino nucleus cross sections. This was soon acknowledged and
a series of workshops (Nuint) fully dedicated to this topic started in 2001.
Even more interesting for hadron physics is the fact that the very high luminosity of the neutrino beams is starting to provide valuable information about the axial properties of baryons, that would be hard to obtain with electromagnetic probes.
 
These reasons have led to much theoretical work~\cite{nuint} studying the basic processes, 
but still many of the codes used in the experimental analysis apply outdated or incomplete models, or they are extrapolated out of their validity limits. 
For example, in  Ref.~\cite{2007ru} an artificial free parameter that
modifies Pauli-blocking was introduced  to fit the low $q^2$ 
QE data.  In the study of  coherent $\pi$ production of 
 Ref.~\cite{Link:2007mf}, some nuclear parameters were modified to unrealistic values. For instance, 
the Fermi momentum was varied up to values corresponding to almost four times the normal nuclear density. These kind of approaches may fit the data and not affect the most important observables, but still the introduction of wrong and uncontrolled physical assumptions in the models leads to some doubts about the reliability of the results.

In this talk, I will present a few selected results of the ongoing work of our group on QE 
scattering and coherent pion production induced by neutrinos.

\section{Quasielastic neutrino nucleus scattering}
  
  This topic has been widely studied. See e.g.~\cite{Nieves:2004wx,Amaro:2004bs,Leitner:2006ww} and references therein.
It is  well known that the impulse approximation (IA) in a Fermi Gas (FG) model, frequently
used in the analysis of neutrino experiments, fails to describe lepton nucleus 
cross sections. Other nuclear effects like the use of proper nucleon spectral functions and
correlations are clearly needed~\cite{Benhar:2005dj}.
Furthermore, the models should be tested on other processes which measure the nuclear response at similar excitation energies. 

Here, we will discuss some of the results of  Refs.~\cite{Nieves:2004wx,Nieves:2005rq} on the QE $\nu$-nucleus process. This model had been successfully used in the study of photonuclear processes 
like $(\gamma, N)$ or $(\gamma, N\pi)$~\cite{Carrasco:1989vq,Carrasco:1992mg} and electronuclear reactions~\cite{Gil:1997bm}.
In these works, apart from the  above mentioned nuclear effects implemented in a consistent many body framework,
 FSI is described  by a MonteCarlo simulation.
 Many of the calculations use either the plane wave (PW) or the distorted wave impulse approximation (DWIA). Neither of them
is appropriate for inclusive processes like QE. In PW, the strong interaction between the final nucleon and the residual nucleus is fully neglected. In the DWIA only the elastic part of this interaction is properly taken into account, whereas the inelastic processes are accounted for by means of the imaginary part of some optical potential that  reduces the nucleon flux and thus, the cross section. However, in these inelastic processes, the nucleon simply changes energy, charge and/or angle and possibly more nucleons are emitted, still contributing  to the QE cross section.
 
The main effect found for the inclusive $(\nu,l)$ reaction, at the intermediate energies where this model is better suited, is the large reduction of the cross section due to the RPA correlations as compared to the IA descriptions. Additionally, the use of realistic nucleon spectral functions distorts the shape of the QE peak producing a significant broadening. 
Concerning the  processes where a nucleon is detected, the inclusion of the FSI means that, even for light nuclei, the nucleon spectrum is strongly displaced towards low energies. For CC events a significant cross section
for the "wrong" charge nucleon appears. See Ref.~\cite{Nieves:2005rq} for a detailed discussion.

\section{Coherent neutrino induced $\pi$ production}
A  model for $\nu$ induced coherent $\pi$
production has been developed in Refs.~\cite{AlvarezRuso:2007tt,AlvarezRuso:2007it,AlvarezRuso:2007sq}.
This model improves upon previous
calculations~\cite{Kelkar:1996iv,Singh:2006bm} by using a more complete description of the
elementary pion production and a better treatment of the pion FSI.
 The process is dominated by the $\Delta$  excitation, but  other 
background pieces like the cross $\Delta$ and nucleon--pole terms are also considered. 
There are many other contributions, relevant for the reaction in a single nucleon~\cite{Hernandez:2007qq},
but, due to their isovector nature, they cancel for isospin symmetric nuclei and we neglect them.
 The purely nucleonic part of the hadronic current is better known and is written in terms of vector and axial form factors (FF).
The vector FF are extracted from electron scattering data and the axial ones are contrained by PCAC.
The  $N-\Delta$ transition is more complex and PCAC is insufficient to fully determine the axial sector.
 The usually called Adler model, which takes  $ C_3^A=0$ and  $C_4^A = - C_5^A /4 $ was chosen and 
 two different parametrizations for  $C_5^A$ ~\cite{Hernandez:2007qq,Kitagaki:1990vs} 
which reproduce the available pion production data on the studied energy region~\cite{Kitagaki:1990vs,Radecky:1981fn}
were considered.

\begin{figure}[ph]
\centering\includegraphics[width=6.in]{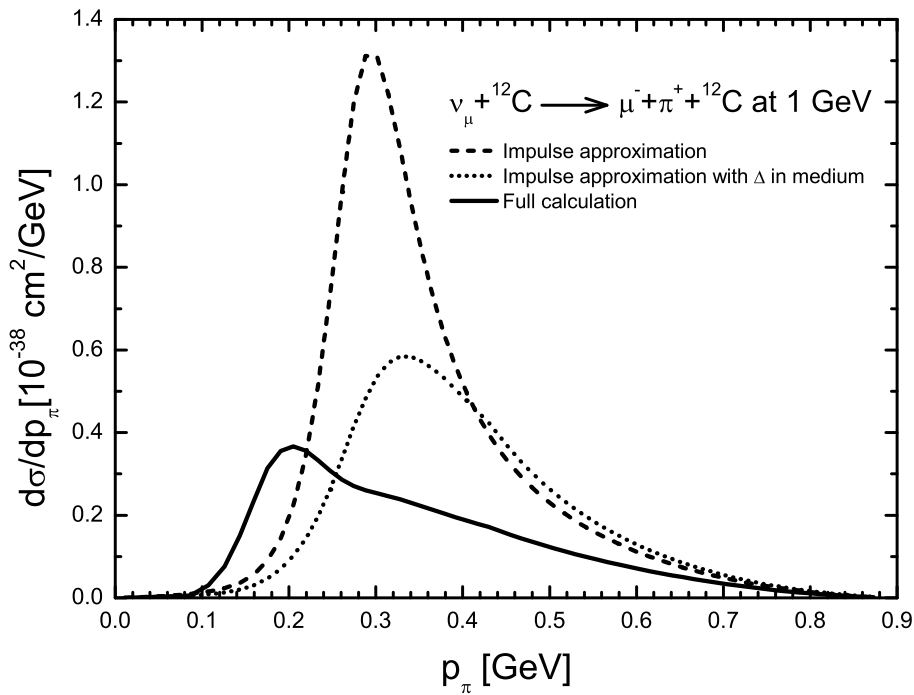}
\caption{Coherent pion production cross section.}
\label{fig1}
\end{figure}
The nuclear medium modification of the $\Delta$ properties is taken into account by modifying its propagator
with the inclusion of an energy and density dependent selfenergy well tested in many $\pi$ nucleus processes.
 This effect, sometimes omitted in other calculations~\cite{Paschos:2005km}, reduces the cross section by around 40~\% 
for 1GeV neutrinos, (see Fig.~\ref{fig1}). There are other nuclear effects like  Fermi motion or Pauli blocking  but the most important one is the FSI of the pion. It is implemented in the DW approximation by solving the Klein-Gordon equation with the optical potential of Refs.~\cite{GarciaRecio:1989xa,Nieves:1991ye}. We find that pion distortion moves the pion spectrum to lower energies and also reduces the cross section. One can notice a large reduction at the $\Delta$ resonance peak due to the large pion absorption at that energy.

The results are more sensitive to $C_5^A$ than in the case of the incoherent reaction. This can be understood from two facts. First, in the forward direction ($q^2 = 0$), dominant in the coherent process, the only FF contributing to the $\Delta$ excitation is $ C_5^A$~\cite{AlvarezRuso:1998hi}. Second, many of the background terms cancel in the coherent sum of amplitudes. Thus, the coherent process is very sensitive to the relatively unknown $N\Delta$ transition axial FF~\cite{Liu:1995bu,Sato:2003rq,Lalakulich:2006sw,BarquillaCano:2007yk,Geng:2008bm} and could eventually provide valuable information complementary to what could be obtained from parity violating electroexcitation of the $\Delta$~\cite{Mukhopadhyay:1998mn}.

\section*{Acknowledgments}
This work was partially supported by MEC contracts 
FIS2005-00810, FIS2006-03438, by the Generalitat Valenciana contract
ACOMP07/302, by Junta de Andaluc\'\i a contract FQM0225 and by the EU 
Integrated Infrastructure Initiative Hadron Physics Project contract
RII3-CT-2004-506078.

\end{document}